\documentstyle[12pt]{article}
\newcommand {\bra}[1]{\langle#1\vert}
\newcommand {\ket}[1]{\vert#1\rangle}
\begin{document}
\centerline{{\huge Two-Electron Quantum Dot in Magnetic Field:}}
\medskip
\centerline{{\huge  Analytical Results}}
\medskip
\centerline{M. Dineykhan and R.G. Nazmitdinov}
\medskip
\centerline{Bogoliubov Laboratory of Theoretical Physics}
\centerline{Joint Institute for Nuclear Research, 141980 Dubna, Russia}
\baselineskip 20pt minus.1pt
\begin{abstract}
Two interacting electrons in a harmonic oscillator
potential under the influence of a perpendicular homogeneous magnetic field
are considered. Analytic expressions are obtained for the
energy  spectrum of the two- and  three-dimensional cases.
Exact conditions for phase transitions due to the electron-electron
interaction in a quantum dot as a function of the dot size and
magnetic field are calculated.
\end{abstract}
\vspace{0.2in}

\vspace{0.2in}
\section{Introduction}

In recent years considerable experimental and
theoretical interest has been focused on the study of artificially structured
atoms in semiconductors, such as quantum dots, where few electrons are
confined in all three directions (see for review
\cite{TC,MK,Jon}).
In experimentally realized quantum dots, the extension in the
$x-y-$plane is much larger than in the $z$ direction.
Assuming that the z-extension could be effectively considered zero,
the electronic properties in these
nanostructures have successfully been described (see \cite{TC,Jon} and
references therein) within the model of the
single-electron motion in the two-dimensional harmonic oscillator
potential in the presence of a magnetic field \cite{Fock}.
Based on a numerical solution of the Coulomb interaction between
electrons, a complex ground state behavior (singlet$\rightarrow$triplet state
transitions) as a function of a magnetic field
has been predicted \cite{MC,Wag} (see also discussion in \cite{MK}).
Remarkably, these ground state
transitions for N=2 have been observed experimentally \cite{As}.

In present paper we consider an analytically solvable model of two
electrons in a quantum dot.
The confinement potential is approximated by a harmonic oscillator potential
and the problem of the Coulomb interaction is treated exactly.
Though this case represents the simplest nontrivial problem with regard
to the electron number, mainly the ground states of a two-dimensional
quantum dot were analysed either quantitatively \cite{Wag} or
analytically with some approximations \cite{Tau}.
The role of third dimension $(z)$ is also investigated and results on
the analysis of the ground as well as excited states are presented.

\section{Model: General Remarks}

Our analysis is based on the oscillator representation method (ORM) developed
in \cite{Din}.
The ORM arisen from the ideas and methods of quantum field theory has been
proposed to calculate the binding energy of different systems with
fairly arbitrary potentials described by the Schr\"{o}dinger equation
\cite{Din1}. Here, for completeness we present briefly the main ideas
of the ORM.

For any potential admitting the existence of a system bound state there is
always a transformation of the variables that leads to a Gaussian
asymptotic form for the wave function at large distances.
However, the asymptotic behavior of the Coulomb wave functions for large
distances does not coincide with this behavior.
Therefore, we have to modify the variables in the original
Schr\"{o}dinger equation so that the modified equation should have solutions
with  the Gaussian asymptotic behavior. In the Coulomb systems, this
modification is performed by going over to the four-dimensional space,
where the wave function of the Coulomb system becomes the oscillator one.
In an early paper \cite{Sc}, Schr\"{o}dinger has noted the existence of
such a transformation which transforms the three-dimensional Coulomb
system into the oscillator one in the four-dimensional space.
The explicit form of this transformation has been found in \cite{Kus}
and used to solve the classical Kepler problem.

In the next stage, it is necessary to represent the canonical variables
(coordinate and momentum) of the Hamiltonian through
the creation and annihilation operators $a^+$ and $a$.
From the Hamiltonian the pure oscillator part with some, yet
unknown, frequency $\omega$ is extracted, i.e.
$H => H_0 + H_{I} = \omega a^+ a + higher~~ order~~ terms$.
The remaining part, i.e.
the interaction Hamiltonian $H_I$, is represented in terms of normal products
over $a^+$ and $a$. In addition, it is required that the interaction
Hamiltonian does not contain terms quadratic in the canonical variables.
This condition is equivalent to the equation
\begin{equation}
\label{cond}
\frac{d\varepsilon_0}{d\omega} = 0
\end{equation}
which determines $\omega$, the oscillator frequency, in the ORM
and is called the {\it oscillator representation condition} (ORC) \cite{Din}.
Similar ideas are used in the Hartee-Fock-Bogoliubov theory to
describe different correlations between nucleons moving in average
nuclear potential (see for review \cite{RS}).

Since we change our space variable $\vec r$ and a magnetic quantum number
$m$ will be absorbed by the dimension parameter of the
auxiliary space (\cite{Din}
and see below), the calculation of the wave function
$\psi(\vec r)$ would be equivalent to the calculation of the ground
state function of a modified Hamiltonian in another dimension.
As it has been mentioned above, the wave functions in this auxiliary space
should have the oscillator Gaussian asymptotic behavior
at large distances. This property is quite natural
for our purposes due to physical conditions for the confined
electron gas in a quantum dot.
The effective confining potential (oscillator) prevents the tendency
caused by the Coulomb forces to allow  electrons to escape and, therefore,
it should dominate in the phenomena.

The model is described by the Hamiltonian
\begin{eqnarray}
H& = &\sum_{j=1}^2{\Bigg\{} {1\over 2  m^{\star}}
(\vec  p_j -{e\over c}\vec  A_j)^2+{m^{\star}\over 2}\left[
\omega _0^2(x_{j}^2+y_{j}^2)+\omega _z^2z_{j}^2 \right] {\Bigg\}}\nonumber\\
     \\
& + &\frac{e^2}{4\pi \epsilon \epsilon_0} \frac{1}{|\vec r_1 - \vec r_2|}
+ H_{spin}\nonumber
\label{ham}
\end{eqnarray}
where $H_{spin} = g(\vec s_1 + \vec s_2) \vec B$.
Here $m^\star$ is the
effective electron mass. Below, we use the units $(e=c=1)$.
For the perpendicular magnetic field
$(\vec B||z)$ we choose the gauge described by the
vector $\vec A =[\vec B\times  \vec r]/2  ={1\over2}\vec B (-y,x,0)$.
Introducing the relative and center-of-mass coordinates
\begin{equation}
\label{sep}
\vec r = \vec r_2 - \vec r_1 \;\;\; \vec R = {1\over2}(\vec
r_1+\vec r_2)
\end{equation}
the Hamiltonian, Eq.(\ref{ham}), can be separated into the
center-of-mass (CM)  and relative-motion (RM)
terms as (see also \cite{Wag,Tau})
\begin{eqnarray}
\label{gen}
H& = &2H_q + {1\over 2}H_Q + H_{spin}\\
\label{rel}
H_q& = &{1\over 2}\left[ {\vec p_q} + {\vec A}_q \right]^2 +
{\hbar^2\over 2} ({\omega_q}^2 \rho_q^2 + \omega_{q_z}^2q_z^2)+
\frac{k \sqrt{\hbar \omega_0}}{2q} \\
\label{cen}
H_Q& = &{1\over2}\left[ {\vec P_Q} + {\vec A}_Q\right]^2 +
{\hbar^2\over 2}({\omega_Q}^2 \rho_Q^2 + \omega_{Q_z}^2 Q_z^2)
\end{eqnarray}
where $\omega_Q = 2\omega_0$, $\omega_{Q_z} = 2\omega_z$,
$\omega_q = {1\over 2}\omega_0$, $\omega_{q_z} = {1\over 2}\omega_z$,
${\vec A}_Q = {\vec A}(q_1) + {\vec A}(q_2)$,
${\vec A}_q = {1\over 2}\left( {\vec A}(q_2) - {\vec A}(q_1)\right)$
and ${\vec A}(q) =\frac{\hbar}{m^\star}[\vec B\times  \vec q]$.
Here we have introduced the variables
$\vec q = \frac{\sqrt{m^\star}}{\hbar}\vec r$,
$\vec Q = \frac{\sqrt{m^\star}}{\hbar}\vec R$,
 $\rho = \sqrt{x^2+y^2}$ and defined
the characteristic lengths: the effective radius
$a^{\star} = a_B \epsilon\frac{ m_e}{m^\star}
 (a_B = 4\pi \epsilon_0 \frac{{\hbar}^2}{m_e e^2})$ and
the oscillator length $l_0 =  (\hbar/m^\star \omega_0)^{1/2}$.
These units allow one to define the dimensionless dot size
$k = l_0/a^\star$ \cite{Wag}.

The separability and the conservation of the angular momentum
lead to a natural ansatz for the eigenfunction of the
Hamiltonian, Eq.(\ref{gen}),
\begin{equation}
\Psi = \psi(\vec q)\phi(\vec Q)\chi(\vec s_1, \vec  s_2),
\end{equation}
where the wave functions $\psi(\vec q)$ and $\phi(\vec Q)$ are \cite{LL}
\begin{equation}
\psi(\vec a) = \frac{e^{im\phi }}{\sqrt{2 \pi}}\psi_m (\rho_a, z)
\end{equation}
and the eigenvalues have the form
\begin{equation}
\label{sum}
E=2\epsilon_r + {1\over 2}E_{N,M}+E_{spin}.
\end{equation}
Here $\epsilon_r$ and $E_{N,M}$ are the eigenvalues of the
Hamiltonians $H_q$ and $H_Q$, respectively. According to the Pauli
principle, if the spatial part of the total wave function is
symmetric (antisymmetric) with respect to the inversion
$r\rightarrow -r$, $\chi$ must be the singlet (triplet)
spin state.

We now concentrate our analysis on the relative motion Hamiltonian $H_r$.

\section{Coulomb problem}

Due to the axial symmetry of the problem,
the Schr\"{o}dinger equation with the Hamiltonian, Eq.(\ref{rel}),
can be written in the form
\begin{eqnarray}
\label{Sr}
{\Bigg\{}-{1\over 2}\left[\frac{d^2}{d{\rho}_q^2} + {1\over \rho_q}
\frac{d}{d\rho_q} - \frac{m^2}{{\rho}_q^2}\right] -
{1\over 2}\frac{d^2}{dq_z^2} +
{\hbar^2\over 2}({\Omega_q}^2 \rho_q^2 + \omega_{q_z}^2q_z^2) \nonumber\\
+\frac{k\sqrt{\hbar\omega_0}}{2\sqrt{\rho_q^2 + q_z^2}}{\Bigg\}}
\psi_m (\rho_q, q_z) = U_m \psi_m (\rho_q, q_z)
\end{eqnarray}
Here
\begin{equation}
\label{Om}
\Omega_q = \omega_q \sqrt{1 + {t^2\over 4}}
\end{equation}
where $t = \omega_c/{\omega_0}$,
$\omega_c=\frac{B}{m^\star}$ is the cyclotron frequency,
and
\begin{equation}
U_m = \epsilon_r - m {{\hbar\omega_c}\over 4}
\end{equation}
where $m$ is a magnetic quantum number.

According to the ansatz of the ORM for the wave function,
we have to change radial variables so as to obtain an oscillator asymptotic
behavior for the wave functions of the transformed equation and
then identify this equation with the radial Schr\"{o}dinger
equation in a space with different dimension.
In addition, our Hamiltonian contains a repulsive
centrifugal term, and the wave function $\psi_m (\rho_q, q_z)$
must decrease at small distances. Consequently, the transformation
to the higher dimensional space is realized by \cite{Din}
\begin{equation}
\label{wf}
\psi_m (\rho_q, q_z) = \rho_q^{d/2 - 1} \Phi_m (\rho_q, q_z)
\end{equation}
The parameter $d$ can be chosen to compensate completely the repulsion at
small distances.
The calculation of the wave function $\Phi_m (\rho_q, q_z)$ is equivalent
to the calculation of the ground state wave functions in a space $R^d$.
The wave function, Eq.(\ref{wf}), is a regular one
at short distances.
Therefore, our wave function possesses necessary
properties at short and large distances, i.e., it is a Gaussian one as
$\rho \rightarrow \infty$ and goes to zero at $\rho \rightarrow 0$.

According to the definition of the wave-function, Eq.(\ref{wf}),
we can transform Eq.(\ref{Sr}) into the equation
\begin{eqnarray}
\label{Sr1}
{\Bigg\{}-{1\over 2}\left[\frac{d^2}{d{\rho}_q^2} +
{{d-1}\over \rho_q}\frac{d}{d\rho_q}\right] -
{1\over 2}\frac{d^2}{dq_z^2} +
{\hbar^2\over 2}({\Omega_q}^2 \rho_q^2 + \omega_{q_z}^2q_z^2) \nonumber\\
+\frac{k\sqrt{\hbar\omega_0}}{2\sqrt{\rho_q^2 + q_z^2}}{\Bigg\}}
\Phi_m (\rho_q, q_z) = U_m \Phi_m (\rho_q, q_z)
\end{eqnarray}
which can be identified with the equation in space $R^d$ with
\begin{equation}
d=2+2|m|
\end{equation}
One can see that the magnetic quantum number $m$  does not enter into the
Schr\"{o}dinger equation, Eq.(\ref{Sr1}), in the explicit form.
It is absorbed by the "dimension" parameter $d$. This trick allows one
effectively to avoid the problem of calculation of excited states
and to perform calculations of the ground state in the auxiliary space $R^d$.
Therefore, Eq.(\ref{Sr1}) contains the oscillator with the coordinate
$\rho_q\in R^d$ and the other one with the coordinate $q_z\in R^1$,
respectively.

Choosing different (fixed) values of $\omega _z$ allows to study the
dependence of the results on the slab thickness.
The condition $\omega _z\gg \omega _0$ ensures that we have a genuine
two-dimensional problem implying that no particles occupie a quantum mode in
the $z$-direction $(n_z=0)$. From the analysis of the far-infrared
frequencies in the three-dimensional oscillator potential
it follows \cite{HN} that the frequency that just
forbids occupation of a $z$-mode is given by $\omega _z^0\ge \omega _{\perp }
(\sqrt{4N+1}-3)/2$ with $\omega _{\perp }$ being the average of $\omega _x$
and $\omega _y$.

\subsection{Two-dimensional case}

Let us consider the case $z = 0$, i.e., the pure two-dimensional electron gas.
For the case under consideration Eq.(\ref{Sr1}) can be written as
\begin{equation}
\label{h0}
H_{2d}\Phi_m (\rho_q) = U_m \Phi_m (\rho_q)
\end{equation}
where
\begin{equation}
\label{tdim}
H_{2d} = -{1\over 2}\left[\frac{d^2}{d{\rho}_q^2} +
{{d-1}\over \rho_q}\frac{d}{d\rho_q}\right]
+{\hbar^2\over 2}{\Omega_q}^2 \rho_q^2 +
\frac{k\sqrt{\hbar\omega_0}}{2\rho_q}
\end{equation}
Here the wave function $\Phi_m (\rho_q)$ depends only on $\rho_q =
\sqrt{\vec \rho_q^2}$. Therefore, we can identify the operator
\begin{equation}
\frac{d^2}{d{\rho}_q^2} + {{d-1}\over \rho_q}\frac{d}{d\rho_q}
\equiv\Delta_{\rho_q}
\end{equation}
with the Laplacian $\Delta_{\rho_q}$ in auxiliary space $R^d$ if
this operator acts on a function depending on the radius only.
The wave function $\Phi_m (\rho_q)$ in Eq.(\ref{tdim}) can be considered
as a wave function of the ground state satisfying the Schr\"{o}dinger
equation
\begin{equation}
H\Phi_m (\rho_q)=\varepsilon(E)\Phi_m (\rho_q)
\end{equation}
where
\begin{equation}
\label{ham1}
H=\frac{P_{\rho_q}^2}2 + {\hbar^2\over 2}{\Omega_q}^2 \rho_q^2 +
\frac{k\sqrt{\hbar\omega_0}}{2\rho_q} - U_m.
\end{equation}
Taking into account Eq.(\ref{h0}), the desired energy $E$ is
determined by the equation
\begin{equation}
\label{eps}
\varepsilon(E) = 0
\end{equation}
Let us express the canonical variables $\rho$ and $p$ through the
creation and annihilation operators $a^{+}$ and $a$
\begin{eqnarray}
\label{rep1}
\rho_j = {1\over {\sqrt{2\hbar\omega}}}(a_j + a_j^{+}), \;\;\;
j = 1,...,d\nonumber\\
     \\
p_j =-i \sqrt{\frac{\hbar\omega}{2}}(a_j - a_j^{+}), \;\;\;
[a_i , a_j^{+}] = \delta_{ij}\nonumber
\end{eqnarray}
where $\omega$ is a new oscillator frequency which is defined below.
The vacuum state $|0>$ is defined according to the standard rules
\begin{equation}
<0|0> = 1, \;\;\; a_j|0> = 0
\end{equation}
Substituting the representation, Eqs.(\ref{rep1}),
into the definition of the Hamiltonian, Eq.(\ref{ham1}),
after some transformations
\begin{eqnarray}
\frac{P_{\rho_q}^2}2 + {\hbar^2\over 2}{\Omega_q}^2 \rho_q^2 & = &
{1\over 2}\left( P_{\rho_q}^2 + \hbar^2\omega^2 \rho_q^2 \right) +
{\hbar^2\over 2}\left({\Omega_q}^2 - {\omega}^2 \right)\rho_q^2\nonumber\\
& \Rightarrow & \hbar\omega \sum_j a_j^{+} a_j +
\hbar\left({{d\omega}\over 4} +
{d\over 4}\frac{\Omega_q^2}{\omega}\right)
\end{eqnarray}
and
\begin{eqnarray}
{1\over {\rho_q}} =
\int\limits_{-\infty}^{\infty}\frac{d\tau}{\sqrt{\pi}}
e^{-\tau^2 \rho_q^2} =
\int\limits_{-\infty}^{\infty}\frac{d\tau}{\sqrt{\pi}}
\int(\frac{d\eta}{\sqrt{\pi}})^d
e^{-\eta^2}e^{-2i\tau (\rho_q \eta)}\nonumber\\
\Rightarrow \sqrt{\hbar\omega}{\Bigg(}
\frac{\Gamma(\frac{d-1}2)}{\Gamma({d\over 2})} +
\int\limits_{-\infty}^{\infty}\frac{d\tau}{\sqrt{\pi}}
\int(\frac{d\eta}{\sqrt{\pi}})^d
e^{-\eta^2(1+\tau^2)}:e_2^{-2i\tau \sqrt{\hbar\omega}(\rho_q \eta)}:{\Bigg)}
\end{eqnarray}
we obtain
\begin{equation}
H = H_0 + H_I + \varepsilon_0
\end{equation}
where
\begin{eqnarray}
H_0& = &\hbar\omega \sum_j a_j^{+} a_j\\
\label{gr}
\varepsilon_0& = &\hbar\left(
{d\over 4}\omega + {d\over 4}\frac{\Omega_q^2}{\omega}\right)
- U_m + {\hbar\over 2}\sqrt{\omega\omega_0}k
\frac{\Gamma(\frac{d-1}2)}{\Gamma({d\over 2})}\\
H_I& = &{\hbar\over 2} \sqrt{\omega\omega_0} k h_I\\
h_I& = &\int\limits_{-\infty}^{\infty}\frac{d\tau}{\sqrt{\pi}}
\int(\frac{d\eta}{\sqrt{\pi}})^d
e^{-\eta^2(1+\tau^2)}:e_2^{-2i\tau \sqrt{\hbar\omega}(\rho_q \eta)}:
\end{eqnarray}
Here $:...:$ means a normal product, and we have introduced
the notation $e_2^x = e^x - 1 - x - {1\over 2}x^2$.
According to the ORM \cite{Din}, the interaction
Hamiltonian $H_I$ does not contain terms quadratic in the
canonical variables, i.e. proportional to $:\rho_q^2:$.

The ORC requirement, Eq.(\ref{cond}),
determines the oscillator frequency $\omega$ which is defined in
the following way:
\begin{equation}
\label{omeg}
\omega = x^2 \Omega_q
\end{equation}
The quantity $x$ is determined with the following equation which is derived
from Eqs.(\ref{gr}), (\ref{cond})
\begin{equation}
\label{x}
x^4 + \frac{x^3}
{\sqrt 2} \frac{k}{(1+\frac{t^2}4)^{1/4}}
\frac{\Gamma({1\over 2}+|m|)}{\Gamma(2+|m|)}
-1 = 0.
\end{equation}
It is clear that at zero Coulomb field ($k=0$) $x\equiv x_0=1$
while for $k\neq0$ Eq.(\ref{x})
defines the effective dependence on the Coulomb interaction
of the oscillator frequency ($k<<1$). Considering the quantity $x$
expanded as a Taylor series in the variable $k$ and keeping only first order
terms $x=x_0 + kx_1 +...$ , we obtain,
according to Eqs.(\ref{Om}), (\ref{omeg}) and (\ref{x})
\begin{equation}
\omega = (1+kx_1)\omega_q\sqrt{1 + {t^2\over 4}} =
{{\tilde \omega}_q}\sqrt{1 + {t^2\over 4}}
\end{equation}
where
\begin{equation}
\label{omq}
{{\tilde \omega}_q} = \omega_q\cdot\left(1-\frac{1}{2\sqrt 2}
\frac{l/a^{\star}}{(1+{1\over 4}t^2)^{1/4}}
\frac{\Gamma({1\over 2}+|m|)}{\Gamma(2+|m|)}\right)
\end{equation}
When the Coloumb forces are absent ($l/a^{\star}=0$) it follows
that ${\tilde{\omega}_q} = \omega_q $ and $\omega = \Omega _q$.

According to the ORM, the quantum number $n$ defines the radial
excitation (see \cite{Din}), i.e., the highest oscillator states
\begin{eqnarray}
\label{vec}
|n> = C_n(a_j^+ a_j^+)^n|0> \nonumber\\
C_n = \left[\frac{\Gamma ({d\over 2})}
{2^{2n} n!\Gamma({d\over 2}+n)}\right]^{1/2}
\end{eqnarray}
Correspondingly, the energy spectrum with radial excitations is defined as
\begin{equation}
\label{rad}
\epsilon^{[n]} (U) \equiv  <n|H|n> = \alpha_1 + \alpha_2
\end{equation}
with
\begin{eqnarray}
\alpha_1& = &({d\over 4} + 2n)\hbar\omega + {d\over 4}
\frac{\hbar\Omega_q^2}{\omega} - U_m\\
\alpha_2& = &{\hbar\over 2}\sqrt{\omega\omega_0}k\left(<n|h_I|n> +
\frac{\Gamma(\frac{d-1}2)}{\Gamma({d\over 2})}\right)\nonumber
\end{eqnarray}
Taking into account Eqs.(\ref{Om}), (\ref{x}), (\ref{rad}) from
Eq.(\ref{eps}) we obtain
\begin{eqnarray}
\label{er}
E_{nm}^{2d}& = & 2\epsilon_r = {\varepsilon}_{nm}^0 +
{\varepsilon}_{nm}^c\nonumber\\
{\varepsilon}_{nm}^0 & = & \frac{\hbar {\omega}_0}{2}
\left[ {m\over 2}t +
(1+|m|+2n)x^2(1+\frac{t^2}4)^{1/2}\right]\\
{\varepsilon}_{nm}^c & = &
\frac{\hbar {\omega}_0}{2} \frac{x k}{2\sqrt{2}}
(1+\frac{t^2}4)^{1/4}\left[ 3
\frac{\Gamma({1\over 2}+|m|)}{\Gamma(1+|m|)}+2<n|h_I|n>\right]\nonumber
\end{eqnarray}
where the matrix $<n|h_I|n>$ is defined in Appendix A.
In perturbation theory the effect of Coulomb forces is taken into account
by the second term ${\varepsilon}_{nm}^c$. In our approach
the main term ${\varepsilon}_{nm}^0 $  depends on the Coulomb forces as well.
In standard schemes this term corresponds to the noninteracting electrons
moving in the external confining potential \cite{Tau}.
Here, within our model, the interaction modifies the external potential
and results in the effective mean field potential of the relative motion.

\subsection{Three-dimensional case}

Despite that $\omega _z\gg \omega _0$,
in real samples the effect of the third direction should be taken
into account, and the prediction based on the pure two-dimensional case
is expected to be modified.

Taking into account the definition Eq.(\ref{tdim}),
Eq.(\ref{Sr1}) can be written in the following form:
\begin{equation}
\label{Sr2}
\left[\left(H_{2d} - U_m\right) + h_z +
h_{res}\right]\Phi_m (\rho_q, q_z) =  0
\end{equation}
where
\begin{eqnarray}
\label{z}
h_z& = &-{1\over 2}\frac{d^2}{dq_z^2}
+ {\hbar^2\over 2}\omega_{q_z}^2q_z^2\\
\label{res}
h_{res}& = &\frac{k\sqrt{\hbar\omega_0}}{2}\left({1\over {\sqrt{\rho_q^2 + q_z^2}}}
- {1\over {\sqrt{\rho_q^2}}}\right)
\end{eqnarray}
Since the terms $H_{2d}$ and $h_z$ give the main contribution to the
total Hamiltonian, the term $h_{res}$ is related to a dimension of
the problem and can be considered as a correction term.
Let us introduce a transformation for the one-dimensional
oscillator $h_z$ similar to the two-dimensional case (see Eq.(\ref{rep1}))
\begin{equation}
q_z = {1\over {\sqrt{2\hbar\omega_z}}}(A^+ + A), \;\;\;
p_z = i\sqrt{\frac{\hbar\omega_z}{2}}(A^+ - A)
\end{equation}
After some transformation of the Hamiltonian, Eq.(\ref{Sr2}), we obtain
\begin{equation}
\label{three}
H = H_0 + \varepsilon_0 + {\hbar\over 2}\sqrt{\omega \omega_0}kh_I
\end{equation}
Here
\begin{eqnarray}
H_0& = &\hbar\omega_{q_z} A^+ A + \hbar\omega \sum_j a_j^+ a_j\\
\varepsilon_0 (U)& = & \hbar\left({d\over 4}\omega +
{d\over 4}\frac{\Omega_q^2}{\omega}  + {{\omega_z}\over 2}\right)
 - U_m\nonumber\\
& + & {\hbar\over {2\sqrt{\pi}}}\sqrt{\omega\omega_0}k
\int\limits_{-\infty}^{\infty}{d\tau}
(1+\tau^2)^{-d/2}(1+\gamma \tau^2)^{-1/2}
\end{eqnarray}
and $h_I$ consists of four terms (see Appendix B). Here $\omega$ is
defined by Eq.(\ref{omeg}) and
$\gamma = \frac{\omega}{\omega_{q_z}}<<1 $.
Finally, applying the definition of the radial excitations
(see Eq.(\ref{rad})) for the three-dimensional case, Eq.(\ref{three}),
from the condition, Eq.(\ref{eps}), we obtain the following
expression for the lowest energy level with $n_z = 0$
\begin{equation}
\label{er1}
E_{nm0} = E_{nm}^{2d} + \frac{\hbar x k}{2\sqrt{2}}
\left[(1+\frac{t^2}4)^{1/4}Q(\gamma) +
2 \frac{\Gamma({1\over 2}+|m|)}{\Gamma(1+|m|)}S_n(\gamma)\right]
\end{equation}
where the quantities $Q(\gamma)$, $S_n(\gamma)$ are defined in
Appendix B.

\section{Discussion}

The solution to the Hamiltonian of the center-of-mass
motion $H_Q$ is well known \cite{Fock} and
the energy can be written as
\begin{equation}
\label{ec}
E_{N,M} =2\hbar {{\omega}_0} \left[ (2N + |M| + 1)\sqrt{1+{t^2\over 4}}
 + (2n_z +{1\over 2}) \frac{\omega_z}{2{{\omega}_0}}
+{1\over 2}Mt\right]
\end{equation}
where $N=0,1,...$ and $M=0,\pm 1,...$ are radial and azimuthal
quantum numbers, respectively.
The spin of the two electrons leads to an additional Zeeman energy
\begin{equation}
\label{es}
E_S= g^\star \mu_B S_z =
{1\over 4}\left[ 1-(-1)^m \right] g^\star
\frac{m^\star}{m_e}\frac{\omega_c}{{\omega}_0}
\hbar {{\omega}_0}
\end{equation}
$m$ is a magnetic quantum number corresponding to the relative
motion and $g^\star$ is an effective Lande factor.

Summing Eqs.(\ref{ec}), (\ref{es}), (\ref{er})
(or Eq.(\ref{er1}) in the three-dimensional case, respectively) we are
able to investigate different ground states as a function of
the dot size $k = l_0 /a{^\star}$ and relative strength of the magnetic field
${\omega_c}/{{\omega}_0}$. Since the center-of-mass quantum numbers
$N, M$ and the quantum number $m$ are conserved by the Coulomb interaction,
the ground state has the quantum numbers $N=0$, $M=0$, $n=0$. Comparing
the energy with different $m\leq0$ we can define the ground state energy
for a given dot size $k$ at different strength of a
magnetic field ${\omega_c}/{{\omega}_0}$.

In our calculations, we used the effective mass $m^{\star }=0.067m_e$
of typical quantum dots for GaAs.
In Fig.1 (a) the energy spectra without a contribution of the Coulomb
forces are presented. While without the Coulomb forces the ground state is
always the state with $m=0$, the Coulomb interaction (Fig.1(b)) leads
to a sequence of different ground states $m= -1, -2,...$ which are
an alternating sequence of singlet and triplet states.

The main mechanism, which defines the optimum quantum number $m$
of the ground state, is the interplay between the dot size and the
strength of the magnetic field. This question has nicely been
discussed for the two-dimensional quantum dots in \cite{Wag} (see also
\cite{MK}).
Similar behavior is observed for the radial excitations with $n=1,2,..$.

If the third extension (z) is taken into account,
the ground phase transitions are shifted to a higher magnetic field
(see Fig.2).
Since the extension of the slab is inversely proportional to the
confining frequency $(\omega_z \sim {1\over {d_z}})$, the thicker the slab
the larger value of the magnetic field is needed
to observe the ground state transition $m\rightarrow m\prime$.
This fact has to be taken into account in experiments.

The singlet-triplet ground phase transition occurs when the following
condition is fulfilled  $E_{0,m} = E_{0,m-1} (m\leq0)$.
For a negative Lande factor the spin-splitting energy in
a magnetic field will lower the energy of the spin $S_z = +1$
component of the triplet states. In particular, the relation
$E_{0, m}=E_{0, m-1}= E_{0, m-2}$ (m odd) defines the point when
the singlet phase ceases to exist \cite{Wag}. Beyond this point
we can observe phase transitions between triplet states
defined by the condition
$E_{0,m} = E_{0,m-2}$ (m odd). Therefore,
at strong magnetic field $\omega_c >>\omega_0$, i.e. in the limit
$t \rightarrow \infty$ and $x\rightarrow 1$, for singlet-triplet
phase transitions $m\rightarrow m-1$ we obtain
\begin{eqnarray}
\frac{l_0}{a^\star} = {8\over 3}
\frac{\Gamma(2+|m|)}{\Gamma({1\over 2}+|m|)}
\left[ (\frac{{\omega}_0}{\omega_c})^{3/2} +
{1\over 2}(\frac{\omega_c}{{\omega}_0})^{1/2}(-1)^m g^\star
\frac{m^\star}m_e \right]f_{st}(\gamma)\\
f_{st}(\gamma) = 1 + {\gamma\over 6}\frac{5+2|m|}{(3-2|m|)(1-2|m|)}
+ O(\gamma^2)
\end{eqnarray}
and for triplet-triplet phase transitions $m\rightarrow m-2$ (m odd)
\begin{eqnarray}
\frac{l_0}{a^\star} = {8\over 3}(\frac{{\omega}_0}{\omega_c})^{3/2}
\frac{\Gamma(3+|m|)}{\Gamma({1\over 2}+|m|)}\frac{4}{(5+4|m|)}f_{tt}(\gamma)\\
f_{tt}(\gamma)= 1 - {\gamma\over 2}\frac{7+6|m|}{(5+4|m|)(3-2|m|)(1-2|m|)}
+ O(\gamma^2)
\end{eqnarray}
In these expressions the pure two-dimensional case is realized in the
limit $\gamma \rightarrow 0$.

The higher excitations in the two-electron quantum dots the lesser
the influence of the Coulomb forces on the "crossing" of levels.
For example, the value of the parameter $k=l_0 /a^{\star}$ for
singlet-triplet phase transitions decreases
with increasing radial quantum number $n$.
In particular, for the two-dimensional system we have obtained the
following relation between
parameters $k=l_0 /a^{\star}$ for a singlet-triplet transition at
different $n$
\begin{equation}
\frac{(l_0 /a^{\star})_{n=1}}{(l_0 /a^{\star})_{n=0}} = \frac{2+|m|}{7+|m|}
\end{equation}
While  the interplay between the magnetic field
and the Coulomb forces determines the features of a phase transition
(singlet $\rightarrow$ triplet) for the ground state ($n=0$) \cite{Wag},
mainly the magnetic field
leads to the phase transitions for the high-lying states $n>0$.

The model allows the calculation of the magnetization $M = -dE/dB$.
Since at low energy the magnetization is closely related to the slope
of the ground energy, at $T=0$ K we obtain for $n=0$
\begin{eqnarray}
\mu & =  & - \frac{dE_{0m}}{dB} = {{\hbar}\over 2}
\Bigg[ m +\frac{x^2(|m|+1)}{2}\frac{t}{\sqrt{1+{1\over 4}t^2}}\nonumber\\
& + & \frac{xk}{2\sqrt{2}}\frac{t}{(1+{1\over 4}t^2)^{3/4}}
\frac{\Gamma(|m|+{1\over 2})}{\Gamma(|m|+1)}f_{\mu}(\gamma)
+ {{g^{\star}}\over {2m_e}}(1- (-1)^m)\Bigg]\\
f_{\mu}(\gamma)& = &1 + {\gamma\over 6}\frac{1}{1-2|m|} + O(\gamma^2)
\end{eqnarray}
As it was mentioned in \cite{MC,Wag}, the phase ground state transitions
would be reflected in sharp discontinuities
in the magnetization. The above exact expression can be useful
for the analysis of the experimental features related to the
phase transitions. Also, it allows one to control the approximation made in
the calculations in \cite{Wag,Oh}.

\section{Summary}

Within the proposed model the analytical expressions for
the energy levels and the magnetization of the two-electron quantum
dots are obtained.
The Coulomb interaction is treated exactly and from the analysis of
the energy spectrum it follows that the interplay
between the Coulomb forces and the magnetic field
are an important ingredients for the prediction of the ground phase
transitions. The Coulomb forces lead to the modification
of the external potential and give rize to the effective confining
potential of the relative motion.
Their contribution in the properties
of single-electron states decreases with the increasing
of the radial quantum number $n$.
Finally, we would like to mention that
the third extension (z) modifies the value of a magnetic filed needed
to observe the phase transition: the thicker slab the larger value of
a magnetic filed. We hope that the results presented here could be useful
for the analysis of the electron properties in two-electron quantum dots
and allow to make a conclusion
on a deviation of the real confining potential from the harmonic
oscillator one.

\appendix
\def\theequation{\thesection.\arabic{equation}}
\section{ Two-dimensional case: matrix $\bra{n}h_I\ket{n}$}
\setcounter{equation}{0}

Here we describe some details of the calculations of the quantity
\begin{equation}
\label{hi}
<n|h_I|n> = \int\limits_{-\infty}^{\infty}\frac{d\tau}{\sqrt{\pi}}
\int(\frac{d\eta}{\sqrt{\pi}})^d
e^{-\eta^2(1+\tau^2)}<n|:e_2^{-2i\tau \sqrt{\hbar\omega}(\rho_q \eta)}:|n>
\end{equation}
Taking into account the following equations
\begin{eqnarray}
\label{exp}
e^{i{\vec k}{\vec a}}e^{i{\vec p}{\vec a}^+}=e^{i{\vec p}{\vec a}^+}
e^{i{\vec k}{\vec a}} e^{-({\vec k}{\vec p})}\nonumber\\
e^{i{\vec k}{\vec a}} {\vec a}^+ e^{-i{\vec k}{\vec a}}=
{\vec a}^+ + i{\vec k}\\
e^{\alpha {\vec a}^+ {\vec a}} {\vec a} e^{-\alpha {\vec a}^+ {\vec a}} =
{\vec a} e^{-\alpha},\nonumber
\end{eqnarray}
the fact that
\begin{eqnarray}
\label{axa}
(a^+ a^+)^n = (-1)^n \frac{d^n}{d{\alpha}^n} e^{-\alpha (a^+ a^+)}
{\Bigg |}_{\alpha=0} =\nonumber\\
(-1)^n \frac{d^n}{d{\alpha}^n}
\int\left(\frac{d\eta}{\sqrt{\pi}}\right)^de^{-\eta^2-2i\sqrt{\alpha}
(a^+\eta)}{\Bigg |}_{\alpha=0}
\end{eqnarray}
and Eq.(\ref{vec}), after some transformation we obtain
\begin{eqnarray}
\bra{n}:e^{-iB(a^+\eta)-iB(a\eta)}_2:\ket{n} = C^2_n
\frac{\partial^{2n}}{\partial\alpha^n\partial\beta^n}\cdot
\sum_{j=2}^{2n}\frac{\left(B^2\eta^2\right)^j}{j!}\cdot
\frac{\left(\alpha+\beta-4\alpha\beta\right)^j}
{\left(1-4\alpha\beta\right)^{j+d/2}}{\Bigg|}_{\alpha,\beta=0}
\nonumber
\end{eqnarray}
where $B=\tau \sqrt{2}$. Using these results,
 we have for Eq.(\ref{hi})
\begin{eqnarray}
\bra{n}h_I\ket{n}= \frac{3}{4}\cdot
\frac{\Gamma(d/2-1/2)}{\Gamma(d/2)}\cdot S_{n}~,
\end{eqnarray}
where
\begin{eqnarray}
\label{4.14}
S_n=\frac{4\Gamma(1+n) }{3\sqrt{\pi}}\cdot\sum_{k=2}^{2n}
\frac{(-1)^k\Gamma(k+1/2)}{\Gamma(k+d/2)}\cdot N_k(n,d)~,\nonumber
\end{eqnarray}
and
\begin{eqnarray}
N_k(n,d)=\sum_{p=0}^{n}\frac{2^{2p-k}\Gamma(k+n-p+d/2)}
{(n-p)!(2p-k)!\left((k-p)!\right)^2}~.
\nonumber
\end{eqnarray}
In a  particular case, $n=1$, and $n=2$  for
$S_{n}$ we have
$$ S_1=\frac{2}{d}~,~~~~~~~
S_2=\frac{4}{d(d+2)}\cdot\left[d+\frac{19}{8}\right]~.$$

\section{Three-dimensional case:
definition of $h_I$, $Q(\gamma)$ and $S_n(\gamma)$}

\setcounter{equation}{0}
Using the same technique as for the two-dimensional case and
omitting tedious calculations, we present the final result
\begin{eqnarray}
h_I& = & h_1 + h_2 + h_3 +h_4\\
h_1& = &\int\limits_{-\infty}^{\infty}\frac{d\tau}{\sqrt{\pi}}
\int(\frac{d\eta}{\sqrt{\pi}})^d
e^{-\eta^2 (1+\tau^2)}
:e_2^{-2i\tau \sqrt{\hbar\omega}(\rho_q \eta)}:\nonumber\\
& & {1\over {\sqrt{1+\gamma \tau^2}}}\left[1
+ \frac{\gamma \tau^2 \hbar\omega_{q_z}}{(1+\gamma \tau^2)}
:q_z^2:\right]\\
h_2& = &\int\limits_{-\infty}^{\infty}\frac{d\tau}{\sqrt{\pi}}
\int\limits_{-\infty}^{\infty}(\frac{dt}{\sqrt{\pi}})
e^{-t^2 (1+\gamma \tau^2)}
:e_2^{-2i\tau \sqrt{\hbar\omega_{q_z}\gamma}(q_z t)}:\nonumber\\
& & {1\over {(1+\tau^2)^{d/2}}}\left[1 + \frac{\tau^2 \hbar\omega}
{(1 + \tau^2)}:\rho_q^2:\right]\\
h_3& = & \gamma \hbar^2 \omega \omega_{q_z}
\int\limits_{-\infty}^{\infty}\frac{d\tau}{\sqrt{\pi}}
\frac{\tau^2}{(1 + \gamma \tau^2)^{3/2}}
\frac{\tau^2}{(1 + \tau^2)^{d/2+1}}:\rho_q^2 : :q_z^2:\\
h_4& = &\int\limits_{-\infty}^{\infty}\frac{dt d\tau}{\sqrt{\pi}}
\int(\frac{d\eta}{\sqrt{\pi}})^d
e^{-\eta^2 (1+\tau^2) - \tau^2 (1 + \gamma \tau^2)}\\
& & :e_2^{-2i\tau \sqrt{\hbar\omega_{q_z}\gamma} (q_z t)}:
:e_2^{-2i\tau \sqrt{\hbar\omega}(\rho_q \eta)}:\nonumber
\end{eqnarray}
\begin{eqnarray}
Q(\gamma)& = &\int\limits_{-\infty}^{\infty}\frac{d\tau}{\sqrt{\pi}}
{1\over {(1+\tau^2)^{d/2}}}
\left[{1\over {\sqrt{1+\gamma \tau^2}}} -1\right]\nonumber\\
& = &- {{\gamma}\over 2}\frac{\Gamma (|m|+{1\over 2})}{\Gamma (|m| + 1)}
+ O(\gamma^2)\\
S_n(\gamma)& = &\frac{\Gamma ({d\over 2})}{\Gamma ({{d-1}\over 2})}
\int\limits_{-\infty}^{\infty}\frac{d\tau}{\sqrt{\pi}}
\int(\frac{d\eta}{\sqrt{\pi}})^d
e^{-\eta^2 (1+\tau^2)}\nonumber\\
& &<n|:e_2^{-2i\tau \sqrt{\hbar\omega} (\rho_q \eta)}:|n>
\left[{1\over {\sqrt{1+\gamma \tau^2}}} -1\right]\nonumber\\
& = &{\gamma \over {2\sqrt{\pi}}}
\frac{\Gamma(n+1) \Gamma(|m|+1)}
{\Gamma(1+|m|+n)}\sum_{l=2}^{2n}(-1)^l
\frac{\Gamma({1\over 2}+l)}
{\Gamma(1+l+|m|)}\\
& &N_l(n,d)\frac{1 + 2l}{1 - 2|m|} + O({\gamma}^2)\nonumber
\end{eqnarray}

\newpage
\centerline{\bf Figure Captions}
{\bf Fig.1} The energy spectrum of a two-dimensional quantum dot in units
of $\hbar {{\omega}_0}$ as a function of the magnetic field strength
${\omega_c}/{{\omega}_0}$. The family of states
with the quantum numbers $N=0$, $M=0$, $n=0$ and $m\leq0$ is shown
(a) without and (b) including the Coulomb interaction between
the two electrons. The arrow indicates the value of the magnetic field
strength ${\omega_c}/{{\omega}_0}= 1.91$ where the second "crossing"
occurs between the lowest states $m=-1$ and $m=-2$.

{\bf Fig.2} Similar to the Fig.1 for three-dimensional quantum dot
$(1/{\gamma} = \omega_z/{{\omega}}=3$) including the Coulomb interaction
between the two electrons. Here, the second "crossing" occurs at
${\omega_c}/{{\omega}}= 3.64$.
The decreasing of the ratio $1/{\gamma}$ leads to the "crossing" of levels at
higher magnetic field strength .

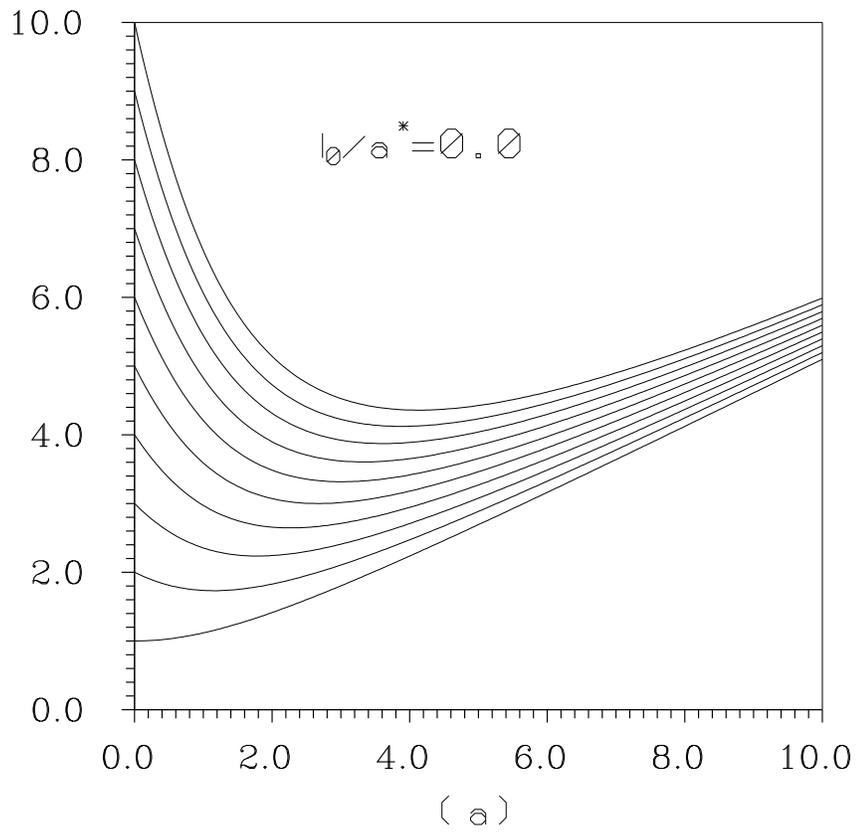
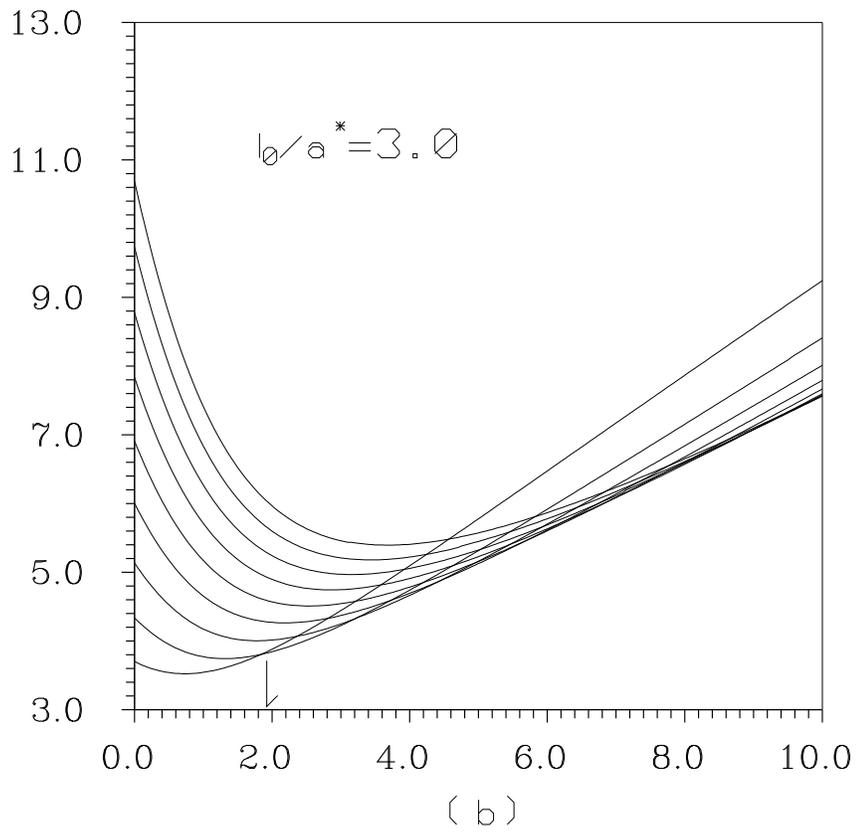
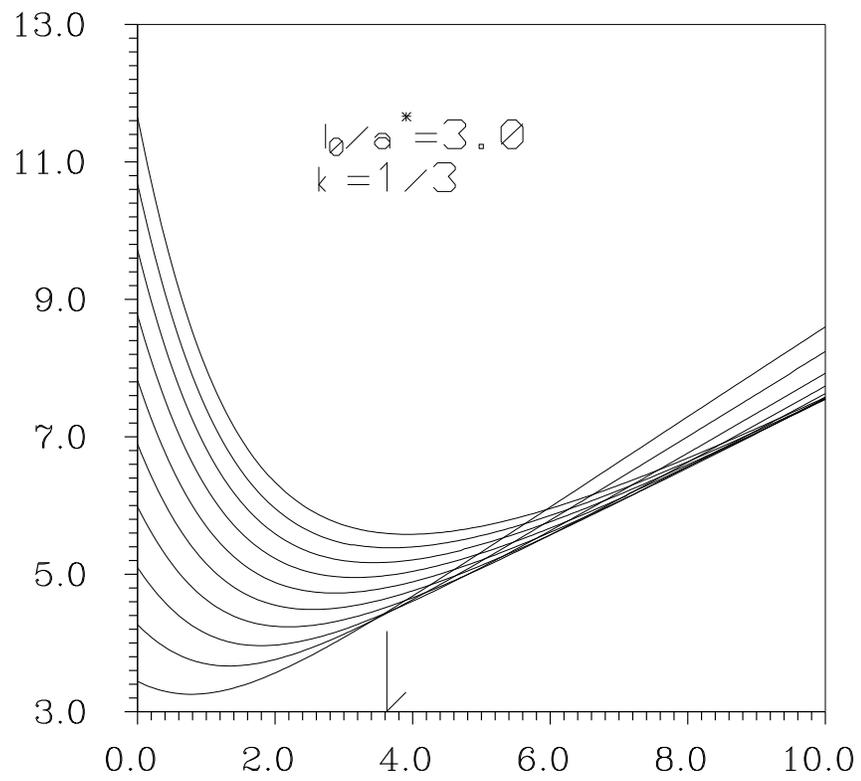
\end{document}